# Multicore fibre technology – the road to multimode photonics

J. Bland-Hawthorn*[a,b,c], Seong-Sik Min [a,b], Emma Lindley [a,b], Sergio Leon-Saval [a,b,c], Simon Ellis [d], Jon Lawrence [d], Nicolas Beyrand [a,e], Martin Roth [f], Hans-Gerd Löhmannsröben [g], Sylvain Veilleux [h]

[a]Sydney Astrophotonic Instrumentation Labs (SAIL), University of Sydney, NSW 2006; [b]Sydney Institute for Astronomy (SIfA), University of Sydney, NSW 2006; [c]Institute of Photonics and Optics (IPOS), University of Sydney, NSW 2006; [d]Australian Astronomical Observatory (AAO), North Ryde, NSW 2152; [e]INSA de Toulouse, 135 Ave de Rangueil, 31400 Toulouse; [f]InnoFSPEC, Leibniz-Institut für Astrophysik, An der Sternwarte 16, D-14482 Potsdam; [g]Institut für Chemie - Physikalische Chemie, Universität Potsdam, Karl-Liebknecht-Strasse 24-25, D-14476 Golm; [h]Department of Astronomy and Space-Science Institute, University of Maryland, College Park, MD

**ABSTRACT**

For the past forty years, optical fibres have found widespread use in ground-based and space-based instruments. In most applications, these fibres are used in conjunction with conventional optics to transport light. But photonics offers a huge range of optical manipulations beyond light transport that were rarely exploited before 2001. The fundamental obstacle to the broader use of photonics is the difficulty of achieving photonic action in a multimode fibre. The first step towards a general solution was the invention of the photonic lantern[1] in 2004 and the delivery of high-efficiency devices (< 1 dB loss) five years on[2]. Multicore fibres (MCF), used in conjunction with lanterns, are now enabling an even bigger leap towards multimode photonics. Until recently, the single-moded cores in MCFs were not sufficiently uniform to achieve telecom (SMF-28) performance. Now that high-quality MCFs have been realized, we turn our attention to printing complex functions (e.g. Bragg gratings for OH suppression) into their $N$ cores. Our first work in this direction used a Mach-Zehnder interferometer (near-field phase mask) but this approach was only adequate for $N$=7 MCFs as measured by the grating uniformity[3]. We have now built a Sagnac interferometer that gives a three-fold increase in the depth of field sufficient to print across $N \geq 127$ cores. We achieved first light this year with our 500mW Sabre FRED laser. These are sophisticated and complex interferometers. We report on our progress to date and summarize our first-year goals which include multimode OH suppression fibres for the Anglo-Australian Telescope/PRAXIS instrument and the Discovery Channel Telescope/MOHSIS instrument under development at the University of Maryland.

*jbh@physics.usyd.edu.au

## 1. INTRODUCTION

The field of astrophotonics emerged in about 2001 with a view to exploring new ways of making observations in optical and infrared astronomy. The field has made real advances in both photonics and instrumentation[4,5]. Examples include beam combination in optical interferometry[6], beam shaping[7], interferometric spectroscopy[8], photonic combs for calibration and metrology[9-11], vortex coronography[12], artificial guide star generation[13,14], microspectrographs[15-17], lantern technology[18-22], photonic reformatting[23-25], fibre Bragg gratings[26,27], Bragg waveguide gratings[28,29], fused hexabundles[30,31], photon angular momentum spectroscopy[32,33], and so forth. In recent years, astrophotonics has given back to telecom with remarkable advances in mode division multiplexing enabled by photonic lantern technology[34-36]. We note that Google Scholar lists *400 references* involving lantern technology.

The telecom industry remains hugely competitive as big corporations compete to find better ways to transfer data around the globe. For decades, we relied on copper wires, then satellites, but today all data (99.999%) move along undersea cables comprising bundled single-mode fibres. Even during an economic downturn, companies fund R&D groups to find better ways to packet, divide and recombine their signals. This has led to a veritable armada of photonic devices as multi-nationals do battle. Complex functions can be printed into a variety of materials – e.g. array waveguides – in order to process light. But the underlying principle remains the same to this day – in order for a coherent action to operate with maximum efficiency, the light must propagate in the fundamental mode ($LP_{01}$). While this mode supports two polarization states, we concentrate on unpolarized modes of propagation. While single-mode fibres can be made to conserve the state of polarization, this is rarely utilized by telecom or any other field. Coupling light into single-mode fibres is notoriously difficult[37] which is an inconvenient truth for astronomers who want to couple as much light as possible into a fibre or a waveguide. Large-aperture fibres typical of astronomy allow light to propagate in many unpolarized modes ($LP_{11}$, $LP_{20}$, $LP_{21}$…) which leads us to the key reason why astronomers have been slow to adopt photonic techniques in their instruments - photonic devices are designed to operate efficiently in one (usually fundamental) spatial mode. In key respects, the invention of the lantern is a major step towards a general solution.

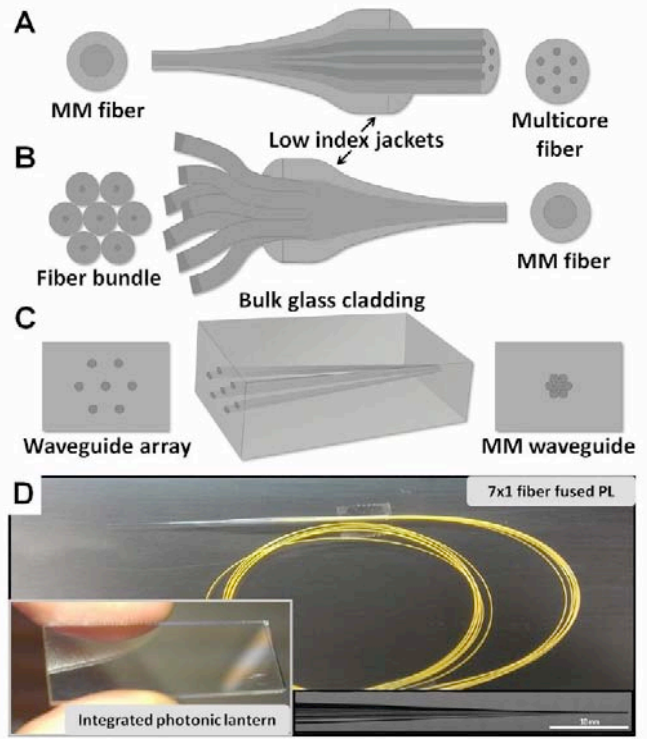

The photonic lantern, first demonstrated in 2005, features an array of single-mode fibres (SMF) surrounded by a low index layer that is adiabatically tapered down to form a multimode fibre (MMF) on input or output depending on the intended direction (Fig. 2). Efficient coupling is achieved in both directions if the number of (unpolarized) excited modes in the MMF is equal to the number of SMFs in the bundle. Light can couple between the bundle of SMFs and the MMF via a gradual taper transition. If the transition is lossless, then the supermodes (group of the degenerate independent SMF modes) of the SMF bundle evolve into the modes (group of non-degenerate supermodes) of the MMF core, and vice versa (Fig. 2). The second law of thermodynamics does not allow lossless coupling of light from an arbitrarily excited MMF into one SMF, but if the MMF has the same number of degrees of freedom as the SMF bundle, then lossless (adiabatic) coupling becomes possible by conserving the entropy of the system.

The original lanterns have an MM input at one end and loose SM pigtails at the other. This gave one the option of purchasing $N$ identical devices (e.g. notch filters) and installing them in parallel. The processed light can then be passed into another lantern (reversed) and the cleaned light subsequently couples into the MM output. This is how the 266 OH suppression gratings operate in the GNOSIS and PRAXIS spectrographs[38,39]. The downside of this approach is the major expense of printing so many independent gratings and then assembling them into the independent lanterns.

Fig. 1 – different forms of photonic lantern based on (A) multi-core fibre; (B) fused SMFs; (C) ultrafast laser injection in a solid. (D) a photograph of lantern B – inset shows lantern C.

It was clear from the outset that more widespread use of the technology would require a different approach to printing gratings into fibres.

*The problem we are presented with today is how to print OH suppression gratings into MMFs.* The SAIL labs have devoted a great deal of time and resources into solving this problem over the past few years. Our preferred solution is based on MCF lantern technology, i.e. where identical gratings are printed across many parallel SM tracks and both ends of the MCF are drawn into MMFs. We believe that this challenge must be overcome if we are to claim that MM photonics has come of age.

## 2. INTERFEROMETERS FOR PRINTING FBGS

FBGs are created by "inscribing" or writing systematic (periodic or aperiodic) variations in refractive index into the core of a photosensitive fibre – typically Ge-doped or hydrogenated – using an intense ultraviolet (UV) laser source. The refractive index (r.i.) of the core changes with exposure to UV light, with the amount of the r.i. change a function of the intensity and duration of the exposure. The dominant processes used are "interference" and "masking". The preferred method depends on the type of grating to be manufactured.

The interferometric method uses two-beam interference (e.g. Sagnac, Mach-Zehnder) and was first used to generate uniform gratings. Here the UV laser is split into two beams that interfere at the core of the photosensitive fibre. The interfering beams create a periodic intensity distribution along the fibre. The r.i. of the fibre changes according to the intensity of UV light. The interference period of the two beams can be modified by changing the incident angle of the beams with respect to each other. This method allows a quick and easy change of the Bragg wavelength.

The masking method utilises a diffractive optical element called the phase mask (PM) to spatially modulate the UV light. To the eye, this looks like ruled glass. The PM is a surface relief grating used in transmission, analogous to the volume phase holographic grating. The PM is placed between the UV laser source and the photosensitive fibre. The spatial diffraction pattern ("shadow") created by the PM determines directly the grating structure along the fibre.

Specifically, when laser light at an appropriate wavelength (e.g. $\lambda_{UV}$ = 244 nm) passes though the PM from above, a periodic or aperiodic (for chirped gratings) variation of phase changes over 0° to 180° along the PM length through diffraction, resulting in an interference pattern when the light leaves the PM (Fig. 3). The PM is generally UV-grade fused silica with a one-dimensional pattern on one surface etched into the silica with a period $\Lambda_{PM}$ (Fig. 2). Some PMs employ a linear chirp in the pattern. The PM maximizes the first orders (*m* = 1 and *m* = -1) in the diffracted light with each containing approximately 35% of the transmitted power. For an orthogonal input beam, the angles of diffraction passing through the PM follow the formula

$$\sin\theta_m = m\,\lambda_{UV}/\Lambda_{PM}$$

The PM design typically suppresses the zeroth and higher orders to below a few percent of the transmitted power at the intended wavelength.

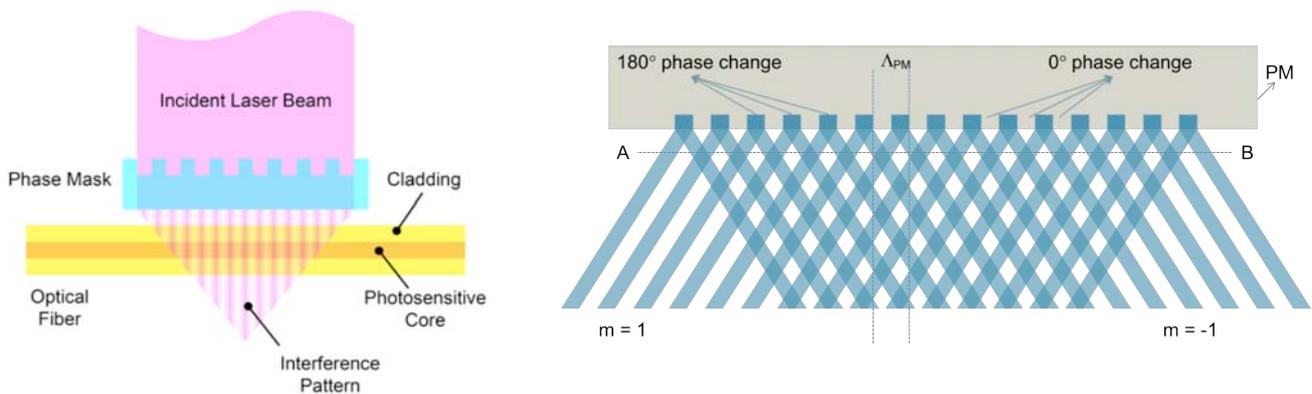

Fig. 2 (left): A phase mask is typically UV-grade fused silica incorporating an etched grating used in transmission (courtesy of Coherent, Inc.). For an orthogonal input beam, the interference pattern has half the period of the phase mask.

Fig. 3 (right): Interference pattern created by the phase mask (PM). The blue and white bands are independent beams so the grating period from A to B is half that of the phase mask. The depth of field can be as much as 100-150 µm which is sufficient to print SMF-28 fibres, but insufficient to print into MCFs. The laser beam arrives at the mask from the top.

Fig. 3 shows a near-field fringe pattern produced by the PM; these are the positive and negative first order diffracted beams. The blue light (as illustrated) has 180° of phase difference from the white light. When the white *m*=1 beams combine with the white *m*=-1 beams, they generate constructive interference; the same holds true for the blue beams. But when the blue *m*=1 light meets the white *m*=-1 light or vice versa, they cause destructive interference. This interference pattern imprints a refractive index modulation in the core of the photosensitive fibre placed in close proximity to the PM;

for example, along the line A to B. The period of the r.i. variation (grating) is half that of the PM as shown in the figure, i.e. $\Lambda = \Lambda_{PM}/2$ where $\Lambda$ is the period of the interference pattern or grating, and $\Lambda_{PM}$ is the period of the phase mask.

The interference pattern becomes weaker as the distance between the fibre and PM increases. A fibre in close proximity to the PM will make stronger gratings in the core. However, if the fibre is placed too close to the PM, it may damage it. If we use $\Lambda_{PM}=1\mu m$, we see that 30 phase changes will participate in making the interference pattern at any instant (Fig. 3), i.e. the beam width in the direction of the fibre is 15μm. In practice, we use beam widths up to 30μm and as small as 4μm to achieve a larger spectral bandpass[27]. In the orthogonal direction, the beam can be *much* wider in order accommodate the width of the fibre (e.g. 500μm or more).

If we set the *z*-axis as the direction from A to B and the *y*-axis as the distance from the PM, the intensity of light or strength of the interference pattern will be

$$I(y,z) = f(y,z)\{u(z-ay) - u(w_O - z - ay)\} e^{-(\alpha y + 2\pi z i/\Lambda)}$$

where *I(y,z)* is the intensity of light, *f(y,z)* is the shaping function corresponding to overall profile of interference pattern, *u(z)* is the step function that confines the interference pattern within the laser beam size, *a* is a reduction constant of interference length according to the distance from the PM, $w_o$ is the laser beam size and *α* is the attenuation constant.

### 2.1 Mach-Zehnder Interferometer (MZI)

The MZI facility in the SAIL labs at the University of Sydney comprises a UV laser system, a beam stabilization system, FBG writing stage, and control systems (Fig. 4). The UV laser system consists of a laser head (1), a remote controller (2), a power transformer/isolator (8), a laser power supply and controller, a heat exchanger (LaserPure 20, Coherent), and a water chiller placed outside of the building (Aqua Cooler).

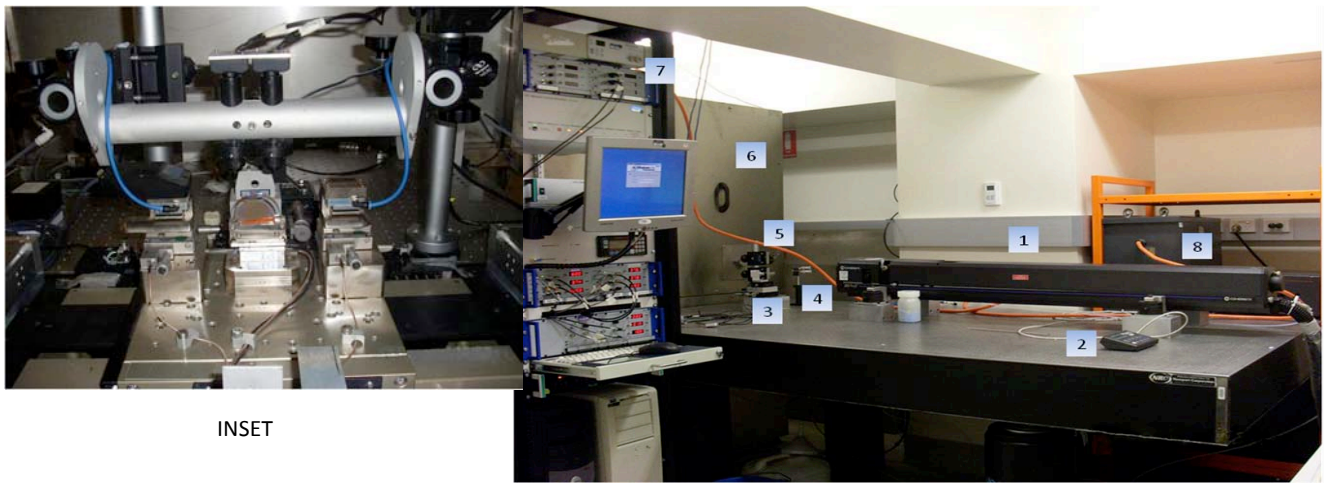

INSET

Fig. 4: SAIL Mach-Zehnder facility: 1 - Argon ion laser, 2 - remote control of laser, 3 - PE controlled mirror, 4 - passive mirror, 5 - height adjust, 6 - FBG writing stage (inset), 7 - main control tower, 8 - laser power supply. The writing stage assembly is illustrated in Fig. 5 where the same numbering is used. The phase mask (PM) is the semi-circular glass element at centre in the inset.

Our MZI laser is an Innova 300 FreD from Coherent: this is an Argon Ion (Ar+) Laser, which generates a laser light of 1 W at maximum (multi-line mode) at 488 nm at the fundamental mode and 100 mW at 244 nm using the frequency-doubling mode. The laser system is controlled by firmware located on the control board inside the power supply. This system is accessed by either the remote control module or the RS-232C interface located on the rear of the power supply. The current to the laser head is 50A at maximum and the efficiency of the laser system is below 0.5% so the laser system generates up to 20 kW of heat and requires a flow of water to cool the head and power supply. The water cooling system consists of a heat exchanger LaserPure 20 (Coherent) and an Aqua Cooler water chiller.

The writing stage is enclosed by a housing (inset in Fig. 4; see Fig. 5) to protect it from dust and stray light. The UV laser beam is guided to the centre of a hole in the writing stage housing (6), as shown in Fig. 4. This is achieved through a piezo-electrically (PE) controlled mirror (3), a passive mirror (4), and a height adjustment mirrors (5). The passive mirror and the height adjustment are responsible for the horizontal and vertical placement of the beam to the hole respectively when the feedback control for the beam stabilization is not used or output values from the feedback controller are zero. This means that the passive mirror is used to set the basic beam path; the PE-controlled mirror stabilizes the beam very sensitively via the feedback controller.

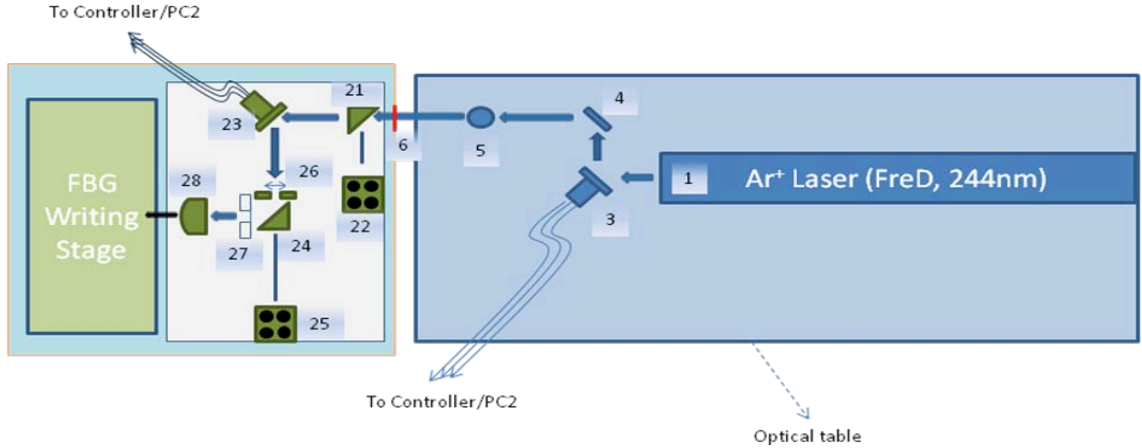

Fig. 5: Laser beam passes from the optical table to the FBG writing stage in the housing. 1: Argon ion laser, 3: Piezo-electrically (PE) controlled mirror, 4: passive mirror, 5: height adjust, 6: hole on the housing for laser beam, 21: beam splitter, 22: quadratic photo-detector, 23: Piezo-electric controlled mirror, 24: beam splitter, 25: quadratic photo-detector, 26: beam shaper, 27: laser shutter, 28: cylindrical lens. The same numbering is used in Fig. 4. The FBG writing stage is illustrated in Fig. 8.

In the writing stage housing, the beam through the hole is directed to the centre of the photosensitive optical fibre, which is placed in close proximity to the phase-mask on the writing stage, using optics shown in Fig. 5, i.e. PE controlled mirror (23), beam splitters (21 and 24), and a cylindrical lens. The cylindrical lens is used to focus the laser beam on to the fibre so the shape of the laser beam at the fibre is elliptical as shown below. In our application, the width of the beam (along the fibre) on the surface of phase-mask is about 30 μm and the orthogonal width is roughly 700 μm. The elliptic beam (Fig. 6b) is achieved using a cylindrical lens before the PM.

**Beam stabilization.** The laser beam is focused onto the PM and the interference pattern generated by the PM is imprinted on the core of the photosensitive fibre (Fig. 2). For an SMF-28 fibre, the core is typically 4 to 10 μm in thickness. The grating reflects light with a specific Bragg wavelength

$$\lambda_B = 2n_{eff}\Lambda_G$$

where $n_{eff}$ is the effective refractive index of the fibre and $\Lambda_G$ is the grating period. For telecom applications, a typical Bragg wavelength is around 1500 nm and the period of grating is near 1 μm with $n_{eff} = 1.5$. Any slight movement or displacement of the laser beam position as small as 50 nm affects the FBG performance either through its reflective or transmissive properties, or both. Since the writing time of an FBG may take more than an hour, beam stabilization is of paramount importance.

In Fig. 5, to implement the beam stabilization, two PE controlled mirrors (3 and 23) and two quadruple photo-detectors, $QPD_1$ (22) and $QPD_2$ (25), are utilized in this system as shown above. The beam stabilization system stabilizes the laser beam from the perturbation on the laser table, the granite table, and writing stage. A small part of laser beam (1%) is sent to the QPD by a beam splitter (21 or 24; Fig. 5) to detect the beam position or displacement. The QPD is composed of four photodetectors (PD) placed on the edges of the detector surface in a shape of a square, as shown in Fig. 5. Each PD is depicted as a dot. The signals from the PDs are transferred to a differential amplifier. This amplifier boosts the amplitude difference between vertical PDs and generates an 'x' signal. The 'x' signal will be a zero if the left and right

PDs generate the same amplitude of signals. It will be positive if the right PD generates a bigger signal than the left and it will increase as the right signal increases and left one decreases. When the left signal is stronger, the 'x' signal will be negative. The amplifier also generates a 'y' signal, which is the vertical difference of signal amplitudes from the PDs and when the upper PDs generate bigger signal, 'y' value will be positive and in opposite case it will be negative. Therefore, if we watch these 'x' and 'y' signals with an oscilloscope when we align the beam path, we can identify the location of the beam on the QPD.

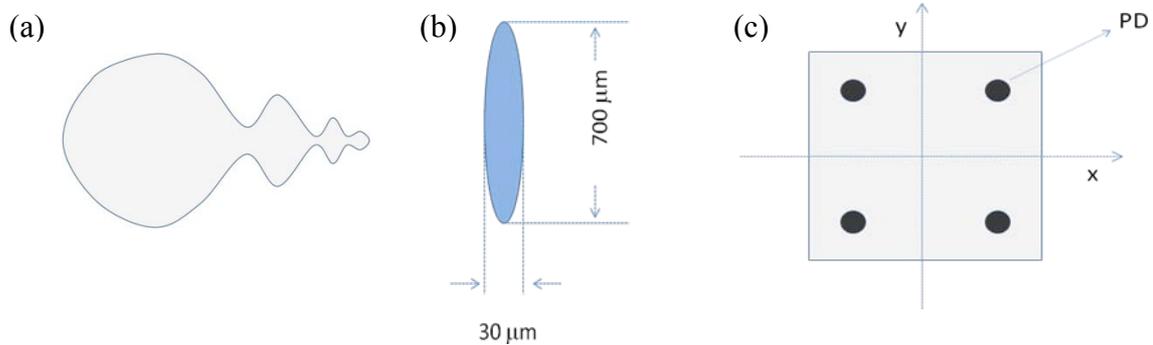

Fig. 6: (a) Actual shape of raw laser beam; (b) laser beam shape on the phase mask after the beam shaper and cylindrical lens; (c) quadruple detector (QPD) made up of four photodetectors (PD).

Similarly, the laser beam coming into the housing is controlled to compensate the perturbation to optical table and the writing stage and to stabilize the laser beam to the optical fibre by two PE mirrors (3 and 23; Fig. 5). The quadruple photo-detectors ($QPD_1$ (22) and $QPD_2$ (25); Fig. 5) are used to detect the beam position on the writing stage. The beam splitters (21 and 24) send 1% of the light to $QPD_1$ first and then $QPD_2$ for the beam alignment.

The QPDs send the signals detected by the 4 PDs to the dedicated difference amplifier. The difference amplifier generates the 'x' and 'y' signals to show the beam position in the detection plane. The 'x' and 'y' signals are sent to the main PLC in the rear of the main control tower (Fig. 7) which is connected to the PC2 with two cables (7E and 7F; Fig. 7). The main PLC sends the 'x' and 'y' signals to the PC2 via these cables by a request. The PC2 reads the signals via a PMAC interface board.

The shape of the raw beam from the I300 FreD laser is complex (Fig. 6a). It has sidelobes which degrade the acuity of the FBG writing. We can remove these side lobes with a beam profiler or shaper by adjusting the distance between the aperture blades and carefully adjusting the 2nd PE mirror manually. During the grating print process, a laser shutter is used to isolate the main circular part and to reject the tail of the laser beam (item 27; Fig. 6). The laser shutter is controlled by the PC1 system and can be opened for testing purposes in real time.

**FBG writing stage.** The writing stage is illustrated in Fig. 8. This consists of outer clamps (16), micro-controllers, a moving stage, a dual micro-camera/monitor, and a red LED to illuminate the PM and fibre for the micro-camera. Inner clamps, a PM holder, a PM, and a dithering device (a PE controlled device) are on the moving stage, which is fully controlled by PC1 via a dedicated power-supply/controller unit made by Aerotech Inc. The PC1 controls everything in the FBG writing stage housing except devices for the beam stabilization described above via the main PLC, simple electrical circuits, and dedicated controllers. The distance between the fibre and the phase mask can be accurately controlled by the micro-positioners (19), a micro-camera (15), the attached monitor (17), and the main control software written in Labview. A phase mask with a period designed for the desired Bragg wavelength is placed in the PM holder (12), which is at the centre of the moving stage (11). The PM holder and PM are placed on a PE controlled device (dithering device) that is fixed to the moving stage.

The photosensitive fibre can be automatically loaded onto the writing stage. The outer clamps hold the fibre and move it towards the inner clamps and the PM. The clamps are stopped by the micro-positioners (19) so that we can actually control the placement of the fibre from the PM by adjusting the position of the micro-positioners. When the fibre is loaded, the system applies some strain to the fibre by pulling apart the outer clamps. The strain sensor below the left outer clamp measures the strain. If the preset strain is achieved, the inner clamps (14) hold the fibre and outer clamps remain open. The inner clamps are fixed on the moving stage by some screws so the fibre and PM can slowly move together left to right or right to left in order to print much longer gratings than the beam width (30 µm) (e.g. travel 40mm at 2mm/min speed). Meanwhile, fast oscillatory movement of the PM can be driven at high frequency over small distances to achieve the "dither function" required in our grating designs (e.g. 0.5um oscillation at 13 Hz).

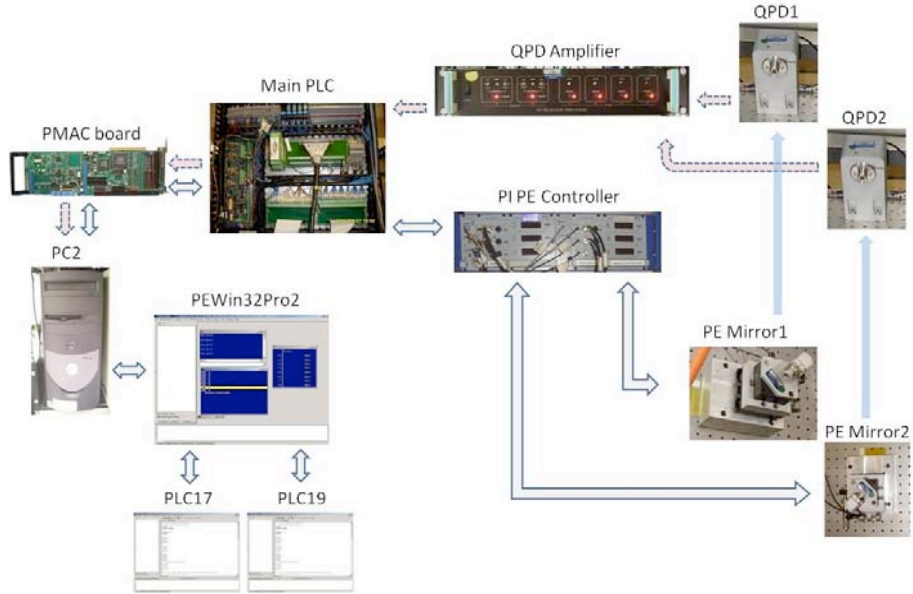

Fig. 7: Control structure for beam stabilization within the MZI interferometer (Figs. 5 and 6).

**Amplitude modulation (apodization) in FBG writing.** The r.i. change increases with the exposure time of the UV light on the photosensitive fibre (Fig. 9a). The higher r.i. produces a stronger grating which induces higher reflectivity in the FBG in the grating. Thus by changing the speed of the moving stage, we can change the degree of reflectivity along the fibre. If the rate of change describes a complex function, we can achieve a sophisticated r.i. amplitude profile for the FBG in order to minimize side lobes, restrict the reflection bandwidth, etc.

The process of achieving a modulated grating amplitude along the fibre length is called apodization. Fig. 9b shows a normalized r.i. change (profile) along the fibre for an FBG with Gaussian apodization. In this figure, the period of grating is about 1 mm for visualization purposes only; the actual period is orders of magnitude smaller. To get the apodization of Fig. 9b, the moving speed of the stage will be much like Fig. 9c. To get the highest r.i. change at the centre, the moving speed is a minimum; to get the zero r.i. change at both ends, the moving speed will be a maximum.

**Dithering.** When the laser passes through the PM, an interference pattern is made as shown in Fig. 3. If we place a photosensitive fibre in the interference region, gratings will be generated in the fibre cores. The PM and the fibre are fixed on the moving stage so if the moving stage travels to left or right, the PM and fibre move together. If the moving stage stops at a position and the laser shutter is open, a small number of gratings within the size of the laser beam spot will be made like in Fig. 10a. If the stage moves to right at a constant speed as shown in Fig. 10b, the PM and the fibre will move too at the same speed because they are fixed on the stage. The laser beam always comes to the same spot and the beam will eventually scan the photosensitive fibre through the PM from right to left and generates longer grating than the beam spot size. The grating length can be up to the length of the PM, i.e. up to 15 cm on our system.

The refractive index of the fibre core at the position of $z$, $n(z)$, can be expressed as

$$n(z) = n_{CO} + a(z)\{1 + \sin(2\pi z/\Lambda_G + \phi(z))\}$$

where $z$ is measured along the fibre axis, $n_{co}$ is the refractive index of the fibre core before the fibre is exposed in the laser light, $\Lambda_G$ is the period of gratings, and $a(z)$ and $\Phi(z)$ are amplitude and phase variations along the fibre length respectively. When $a(z)$ and $\Phi(z)$ are constant, the FBG is called a uniform grating. The grating with non-uniform $a(z)$ is called an *apodized grating* and the grating with non-uniform $\Phi(z)$ is called a *chirped grating*.

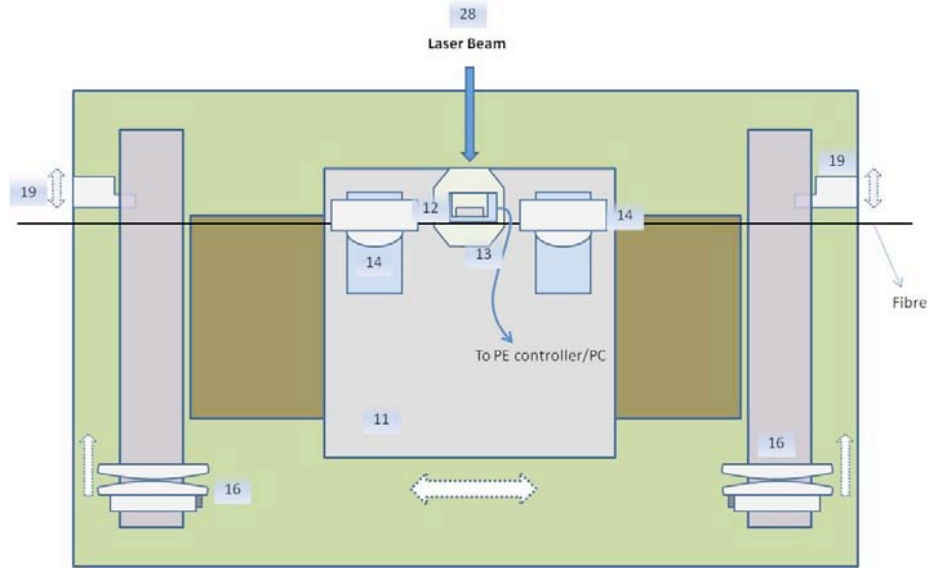

Fig. 8: FBG writing stage. 11: moving stage, 12: phase mask (PM) and holder, 13: Piezo-electric device for dithering, 14: inner clamps, 16: outer clamps, 19: micro-positioners, 28: cylindrical lens.

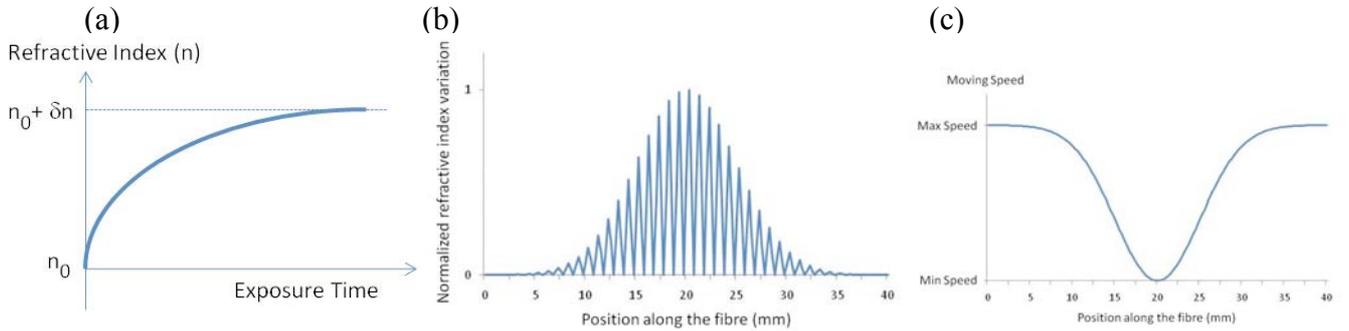

Fig. 9. (a) Exposure time of UV light vs. refractive index change in a photosensitive fibre; (b) Gaussian apodized FBG; (c) Moving speed of writing stage to achieve the Gaussian apodized FBG.

The apodization can be achieved by two methods; one is to use the variation of moving speed of the stage as explained in previous section. The other one is to use the dithering. Dithering is to use the fast periodical movement of the PM within the range of the period of the PM and can be expressed by

$$D(t,z) = O(z) + A(z)\sin(\omega_d t + \vartheta_d(z))$$

where $D(t, z)$ is the dithering at time $t$ and position $z$ along the fibre, and $O(z)$ is the offset value of the dithering. $A(z)$ is the amount of the periodical movement of PM, $\omega_d$ ($=2\pi f_d$) is the basic angular frequency of the dithering, the normal frequency $f_d$ is 10-15 Hz, and $\theta_d$ is the phase change. By dithering, we apply chirp or apodization or both to the FBG.

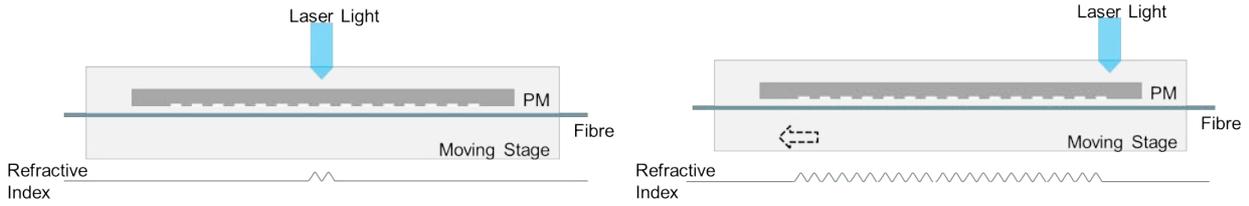

Fig. 10: Basic FBG writing: (a - left) spot FBG, (b - right) uniform FBG. The moving stage + PM + fibre are kept fixed in (a) whereas in (b) the moving stage + PM + fibre have moved to the left.

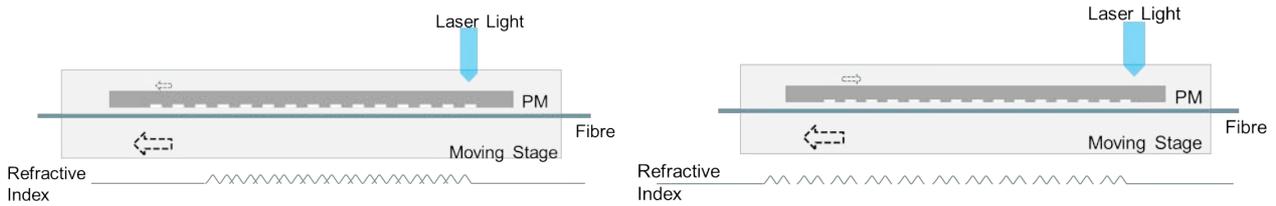

Fig. 11: (a) FBG writing with step movement of the PM. The PM moved (a) in the same or (b) opposite direction as the moving stage.

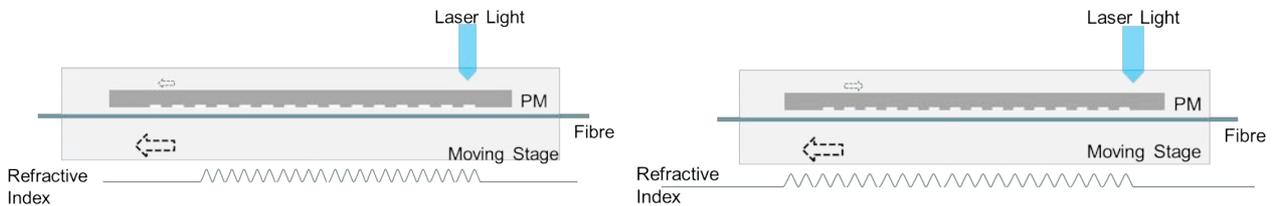

Fig. 12: FBG writing when the PM moved smoothly (a) in the same or (b) opposite direction as the moving stage.

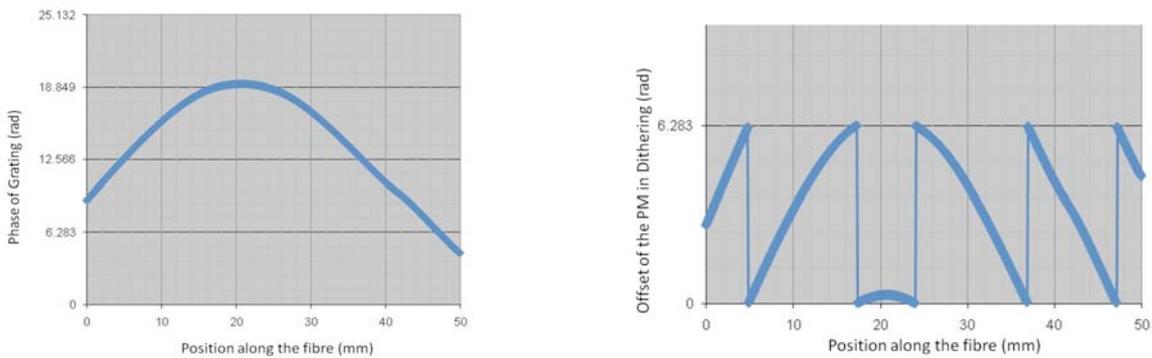

Fig. 13: Relationship between the (a) chirp in the FBG and the (b) offset of the dithering (movement of PM).

To understand the dithering, consider the case of Fig. 11 where the PM moves one step to the right (a) or left (b) while the moving stage (fibre and PM) travels to the right at a constant speed. When the PM moves in the same direction as the moving stage (Fig. 11a), the refractive index inscribed in the fibre core will be overlapped and result in the shortening of the period. In the opposite case where the direction of movement of the PM is reversed (Fig. 11b), the gratings inscribed have gaps in between.

If the PM moves continuously in the same or opposite direction to the moving stage, the generated gratings will be like those shown in Fig. 12. In the case of same direction, the period of the gratings will be shortened (Fig. 12a) and in the case of the opposite direction, the period will be increased (Fig. 12b). This movement of the PM is expressed as the offset $O(z)$ in the above dithering equation.

By controlling this offset (Fig. 13b) along the fibre $O(z)$ we can achieve the chirp (Fig. 13a) because the movement affects the phase of the gratings. The movement is limited to fractions of the period of the PM; the movement of one period is same as no movement assuming that the PM does not have a chirp. In Fig. 13b, the values are in radians but in the actual control of the moving stage, the values are converted to distance in microns or nanometres by the Labview codes.

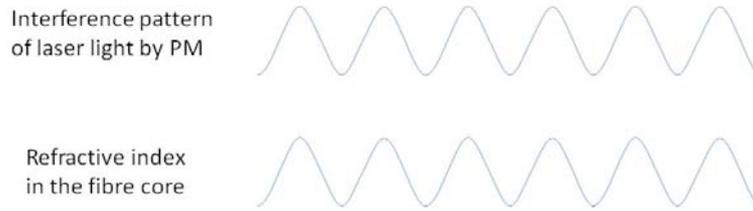

Fig. 14: Amplitude of the simplest grating written by the UV interference pattern (no dither).

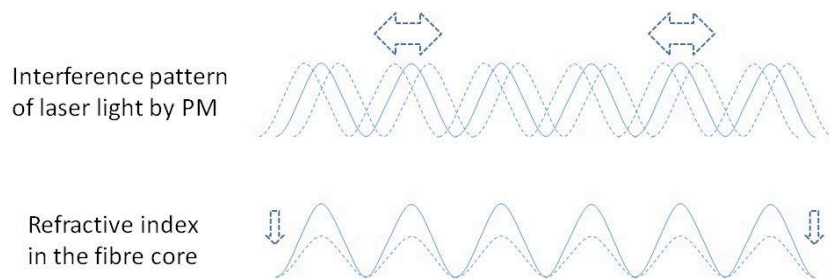

Fig. 15: Amplitude function of the simplest grating with left-right dither; the grating amplitude $a(z)$ is now diluted by the dither.

A key function of the dithering is to control the amplitude or strength of gratings. If the PM is fixed on the moving stage and does not move, the PM generates an interference pattern on the fibre as shown in Fig. 14 (upper) and the interference pattern will be inscribed in the fibre core like Fig. 14 (lower). If the PM moves in a sinusoidal way across the interference pattern (10 to 15 Hz) as shown in the Fig. 15 (upper), the interference pattern will be blurred and the strength of the interference pattern will be weakened so the generated gratings will be weaker as shown in the Fig. 15 (lower). A larger dithering amplitude $A(z)$ within one period results in weaker grating strengths, $a(z)$.

The MZI has been an effective instrument for printing gratings into single mode fibres. We were also able to print nice gratings into a 7-core MCF[3,40]. These results are presented in a separate 2016 SPIE paper, work that is led by Emma Lindley as part of her PhD at the University of Sydney. But our ultimate goal is to print MCF gratings where $N \sim 100$ cores or more. We tried to print such devices but the limitations of our small depth of field were apparent in the final gratings.

**2.2 Sagnac Interferometer**

Last year, we took the bold step of commencing the construction of a new interferometer (Fig. 16) with a view to increasing the depth of field of the UV interference region. The effort and cost behind an interferometer (>$1M) is equivalent to building a highly stabilized, sophisticated astronomical instrument. Essentially *all* aspects of a modern instrument are found in the construction of an interferometer, including control electronics and software, feedback loops, optics, gratings, detectors (for monitoring), engineering, environment control, etc.

The Sagnac layout is more challenging to get right but, once completed, it has *many* advantages. These include (i) the increased depth of field; (ii) the same PM can be used over a broader range of grating designs; (iii) the laser power can be monitored just below the PM, inside the diamond in Fig. 16; (iv) the amplidude of the two beams can be modulated independently by a transmissive acousto-optic device to give extra complexity to the grating -- this is a possible upgrade path; (v) the fibre is now clear of the PM and less susceptible to damage; and so forth. The major disadvantage is the need to control all aspects of alignment to high precision over long periods.

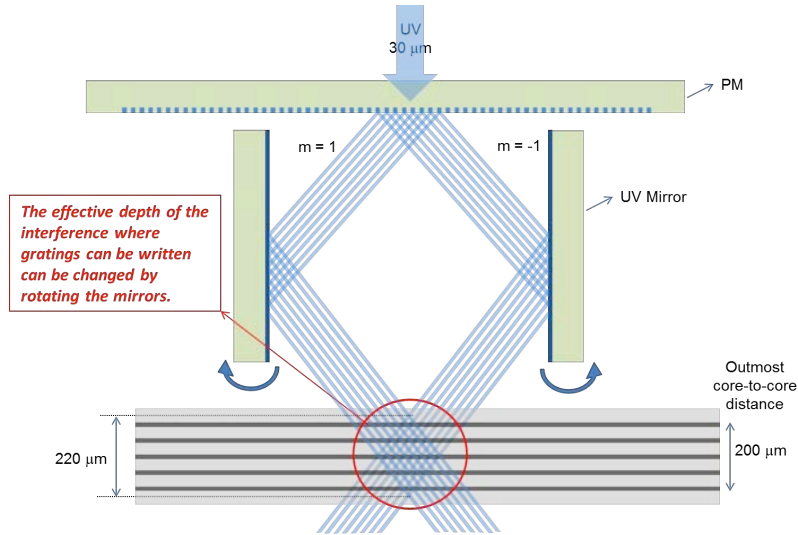

Fig. 16: Schematic of our Sagnac interferometer now in the final stages of construction. The phase mask layout of the MZI is repeated, but the separate orders are now brought together at another location through two mirrors. Once again, the laser beam comes in from the top. Note that the interference depth of field is now twice the depth at the PM; a small tilt of the mirrors can increase the depth of field even further. This design has many advantages over the MZI layout.

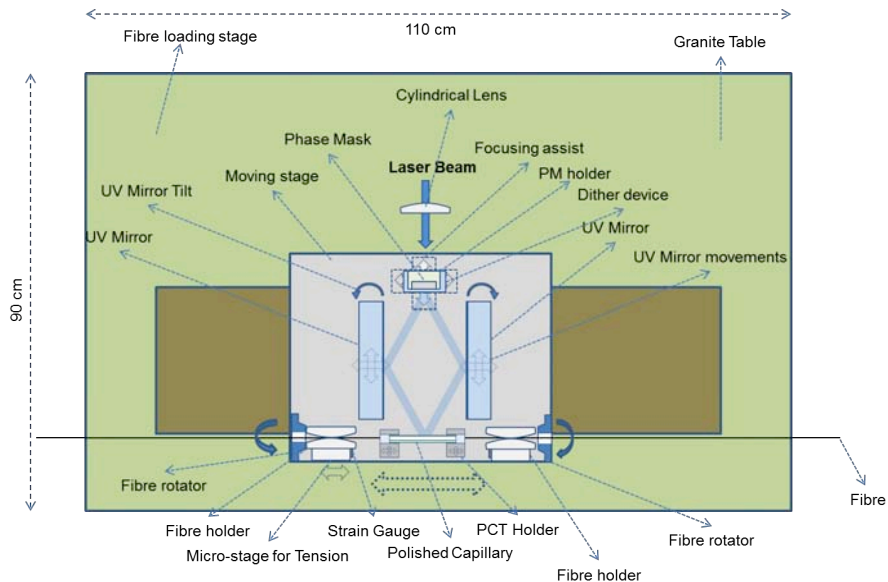

Fig. 17: Schematic of our Sagnac interferometer showing the overall system architecture. Like the MZI (Fig. 8), the laser beam, clamped fibre and PM are kept stationary while the stage moves left or right.

In the vicinity of the FBG writing stage, many aspects of the MZI reappear in the Sagnac layout (Fig. 17). For now, we will be using the dither and scan speed approach to grating writing. But we have the option of including acousto-optic tunable filters (AOTF) in order to modulate the intensity of each beam. The AOTF technique was used to great effect in producing the complex filters presented elsewhere[26,27].

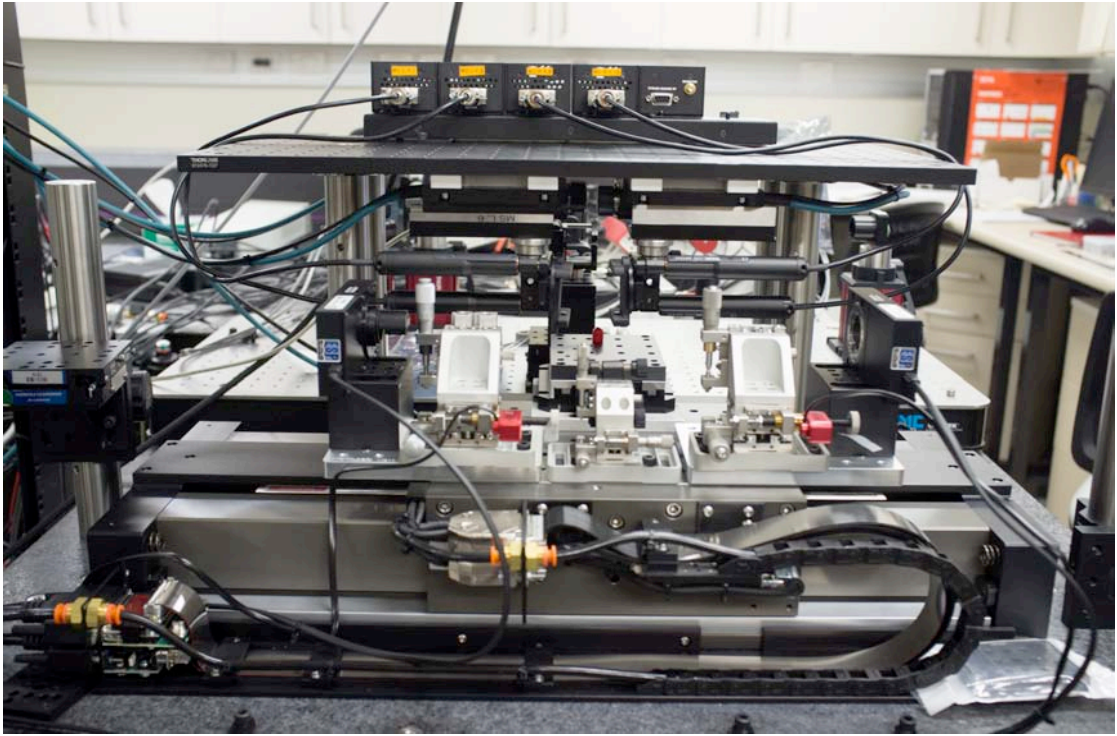

Fig. 18: SAIL Sagnac interferometer under construction. The moving stage is in the foreground. The red spot in the centre of the image is the eventual location of the phase mask. The white blocks show the positions of the multicore fibre clamps. Above the blocks, the horizontal black bars hold the two mirrors that direct the beam after the PM towards the MCF in the clamps. The laser beam emerges from the far field and is directed through the PM out of the page.

## 3. EARLY FBG RESULTS WITH MULTI-CORE FIBRES

Emma Lindley has already presented our first results for a 7-core fibre[3]. These were manufactured with the MZI in the SAIL labs and it was possible to achieve repeatable high-quality results. For a video discussion of this work, we refer the reader to www.jove.com/video/55332 and the matching journal paper[40]. Numerous technical challenges were overcome as explained in the video. We polished one side of a capillary flat to ensure that the laser beam provided flat illumination across the cores. This effect was first observed by a team at the University of Bath[41].

Of the seven cores, the six outer cores provided nice deep 30 dB notches, all in phase. Unfortunately, the central core had a comparable depth but is out of phase by about the notch bandwidth. The outer 6 cores in combination achieve the design depth, but the overall performance is washed out in combination with the middle core. This is an outstanding problem which is either a consequence of the MCF manufacturing process or, more likely, a limitation of our current hydrogenation chamber. The MCFs are hydrogenized to enhance their responsivity to UV light. Tests are now under way to determine the root cause of this problem. If we cannot solve this easily, there are a few post-tuning ideas that we can explore.

## 4. APPLICATION TO NIGHT SKY SUPPRESSION

The night sky problem has been discussed in detail by many authors over the years[42]. With reference to photonic solutions, we have discussed[26] the use of fibre Bragg gratings and demonstrated an 18-notch filter in an SMF. We published[27] the first demonstration in an MMF achieving 105 notches with bandwidth 200pm and 30dB depths over the spectral region 1450-1700 nm. Below, we describe their application in the prototype instrument GNOSIS on the Anglo-Australian Telescope (Section 4.1) before discussing a new instrument PRAXIS that is fully optimized for the MM FBGs (Section 4.2).

## 4.1 GNOSIS

This prototype instrument GNOSIS demonstrated the feasibility of suppressing many OH sky lines (105 lines) in the near infrared. The overall efficiency was very low (3%) because the cleaned light was fed to an existing (not optimally matched) infrared spectrograph. This saved time because we could make use of the spectrograph's infrared detector rather than build a detector system from scratch. Our primary goal was to prove the suppression capabilities of the FBGs. The instrument and its operation has been discussed in detail elsewhere[27,38,39,43].

Here, the OH suppression unit was based on 7 multimode inputs (photonic lanterns), each of which fed $N$=19 FBGs. The cleaned output signal from the FBGs was then sent back through 7 reverse lanterns in order to produce the multimode output that is to be dispersed by the sapphire grism. We have presented the spectacular results at earlier SPIE conferences and in the papers cited above.

## 4.2 PRAXIS

After the success of the prototype, it was clear that to enhance the overall throughput by an order of magnitude, the lanterns would have to be optimally designed to couple efficiently to a high performance IR spectrograph. The instrument was designed and built around a Hawaii 2RG IR detector to be supplied by InnoFspec at the University of Potsdam. A useful summary of progress is given elsewhere[44] with the most recent update provided at this meeting by S.C. Ellis. The expected throughput for the full PRAXIS system is ~30%. Once the MCF FBGs are demonstrated, these will be used in the Maryland OH Suppression Spectrograph (MOHSIS) on the 4.3m Discovery Channel Telescope and ultimately the PRAXIS spectrograph.

JBH acknowledges support from the Australian Research Council through a Laureate Fellowship, LIEF and Discovery Project grants. SV acknowledges partial support for this research through grant AST/ATI 1207785 from the National Science Foundation.